\documentstyle[12pt,aaspp4,flushrt,epsf]{article}

\def\ifm#1{\relax\ifmmode#1\else$\mathsurround=0pt #1$\fi}
\def\kms{\ifmmode\,{\rm km}\,{\rm s}^{-1}\else km$\,$s$^{-1}$\fi}
\def\hmpc{\,h\ifm{^{-1}}{\rm Mpc}}
\def\kmsmpc{\,{\rm km\,s^{-1}Mpc^{-1}}}

\def\m3{Mark\ III}

\def\fig #1, #2, #3 {
  \smallskip
  \centerline{\psfig{figure=#1,height=#2 in,width=#3 in}}
           }
\def\capt{\small \baselineskip 12pt }
\def\be{\begin{equation}}
\def\ee{\end{equation}}

\def\\{\hfill\break}

\def\ra{\rangle}
\def\la{\langle}

\def\sss{\scriptscriptstyle}

\def\etal{{\it et al.\ }}

\def\equ#1{equation~(\ref{eq:#1})}
\def\Fig#1{Figure~\ref{fig:#1}}

\def\se#1{\S~\ref{sec:#1}}

\def\ltsima{$\; \buildrel < \over \sim \;$}
\def\lsim{\lower.5ex\hbox{\ltsima}}
\def\gtsima{$\; \buildrel > \over \sim \;$}
\def\gsim{\lower.5ex\hbox{\gtsima}}

\def\pmb#1{\setbox0=\hbox{#1}%
 \kern-.025em\copy0\kern-\wd0
 \kern.05em\copy0\kern-\wd0
 \kern-.025em\raise.0433em\box0}
\def\vr{\pmb{$r$}}

\def\vv{\pmb{$v$}}

\def\v0{\pmb{$0$}}
\def\vnabla{\pmb{$\nabla$}}
\def\div{\vnabla  \cdot  }

\def\divv{\div\vv}

\begin{document}

\title{THE LARGE-SCALE TIDAL VELOCITY FIELD}

\author{
Y. Hoffman \altaffilmark{1}, A. Eldar \altaffilmark{1},
S. Zaroubi \altaffilmark{2}, \&  A. Dekel \altaffilmark{1}}

\altaffiltext{1}{Racah Institute of Physics, The Hebrew University,
Jerusalem 91904, Israel}
\altaffiltext{2}{Max-Planck-Institut fuer Astrophisik,
Karl-Schwarzschild-Str. 1, 85740 Garching, Germany}

\begin{abstract}

We present a method for decomposing the cosmological velocity field in 
a given volume into its divergent component due to the density 
fluctuations inside the volume, and its tidal component due to the 
matter distribution outside the volume.  The input consists of the 
density and velocity fields that are reconstructed either by POTENT or 
by Wiener Filter from a survey of peculiar velocities.  The tidal 
field is further decomposed into a bulk velocity and a shear field.  
The method is applied here to the \m3\ data within a sphere of radius 
$60\hmpc$ about the Local Group, and to the SFI data for comparison.  
We find that the tidal field contributes about half of the Local-Group 
velocity with respect to the CMB, with the tidal bulk velocity 
pointing to within $\sim 30^{\circ}$ of the CMB dipole.  The 
eigenvector with the largest eigenvalue of the shear tensor is aligned 
with the tidal bulk velocity to within $\sim 40^\circ$.  The tidal 
field thus indicates the important dynamical role of a super attractor 
of mass $(2-5)\times 10^{17} M_\odot h^{-1} \Omega^{0.4}$ at $\sim 150 
\hmpc$, coinciding with the Shapley Concentration.  There is also a 
hint for the dynamical role of two big voids in the Supergalactic 
Plane.  The results are consistent for the two data sets and the two 
methods of reconstruction.

\end{abstract}

\subjectheadings{cosmology: observation --- cosmology: theory ---
dark matter --- galaxies: clustering --- galaxies: distances and redshifts ---
galaxies: formation --- large-scale structure of universe}

%%%%%%%%%%%%%%%%%%%%%%%%%%%
\section{INTRODUCTION}
\label{sec:intro}
%1

%YH The relation between \delta and v in never local. Even when going 
%from v to \delta one has to take a derivative - an act which is not 
%local. Note that in POTENT you have to heavily smooth before taking 
%a derivative. This implies the non-local nature of the relation.
%The
%local nature %YH
%D [it is local when you go from v to density. Poisson is local]    
%YH
The simple relation between the velocity and density fluctuations
in a (linear)  cosmological gravitating system allows us    %YH
to use observed peculiar velocities via their local spatial variations 
to uncover the underlying, mostly dark, mass distribution (reviews: 
Willick 1999 and Dekel 1999, 2000).  This yields interesting 
constraints on the cosmological density parameters, which are free of 
``biasing" --- the unknown relation between galaxy and mass density.  
However, the velocity at a point is determined by the integral over 
the matter distribution in a large volume.  This non-local feature 
enables us to probe the mass distribution in large regions of space 
not directly sampled by peculiar-velocity data.  This is the focus of 
the current paper, along the general lines of the fruitful tradition 
in astronomy to learn about unobserved mass from observed velocities.

The linear theory of gravitational instability provides a simple 
framework for this task.  On scales large enough, where linear theory 
prevails, the linearized continuity and Poisson equations state that 
the mass density fluctuation field is simply proportional to the 
divergence of the peculiar velocity field.  This implies that the 
velocity at a point is given by a spatial integral involving the 
density fluctuation field everywhere about this point.  Given a 
specific volume, this enables a simple decomposition of the full 
velocity field into two components.  The {\it divergent\,} component 
is the part induced by the local mass distribution within the volume, 
and the {\it tidal\,} component is the complementary part induced by 
the mass distribution outside the volume.  In mathematical terms these 
are essentially the ``particular" and ``homogeneous" solutions of the 
Poisson equation.  The tidal field is the component that sheds light 
on structure not directly explored by the velocity survey.

In a pioneering attempt, Lilje, Yahil \& Jones (1986) analyzed a local 
sample of peculiar velocities in our $\sim 30\hmpc$ vicinity (Aaronson 
\etal 1982) to probe the tidal field in the Local Supercluster.  By 
fitting the tidal component to a single-attractor model, they 
predicted the existence of a dominant mass concentration at a distance 
of $\sim 40-50\kms$ outside the Local Supercluster in the direction of 
the Hydra and Centaurus clusters, later detected directly by larger 
samples of peculiar velocities and termed The Great Attractor (GA, 
Dressler \etal 1987).  Here we extend this general approach by 
applying it to a much larger and more accurate sample of peculiar 
velocities, and by using much more advanced tools of reconstruction.

The data used in this paper is the \m3\ catalog of peculiar velocities 
of galaxies and clusters.  The radial velocities are used to recover 
the three-dimensional velocity field and the mass density field either 
by the direct POTENT method (Dekel \etal 1999) or by the Wiener Filter 
(WF) method (Zaroubi, Hoffman \& Dekel 1999).  The uncertainties in 
the recovered tidal field are evaluated by means of mock catalogs 
(POTENT) or constrained realizations (WF).  The volume of reference is 
taken to be the sphere of radius $60\hmpc$ about the Local Group, 
encompassing the local giants --- the Great Attractor and Perseus 
Pisces (e.g., see maps in Dekel \etal 1999).

This paper is organized as follows.  In \se{recon} we briefly describe 
the POTENT and WF methods used to recover the velocity and 
mass-density fields and their corresponding error estimations.  In 
\se{decomp} we describe the decomposition procedure, apply it to the 
\m3\ velocity field, and address the divergent component.  \se{tidal} 
focuses on the tidal component of the velocity field, and its analysis 
in terms of a bulk velocity and a shear tensor.  A simplified 
interpretation of the tidal field using a toy model of a single 
attractor is presented in \se{shapley}.  A general summary and 
discussion is given in \se{conc}.

%%%%%%%%%%%%%%%%%%%%%%%%%%%%%%%%%%%%%%%%%%%%%%%%%
\section{POTENT AND WIENER RECONSTRUCTIONS}
\label{sec:recon}
%2

We reconstruct the full velocity and density fields from the \m3\ radial
peculiar velocities via two alternative methods, POTENT and WF.
The \m3\ catalog (Willick \etal 1997) contains $\sim\!3000$ galaxies
with Tully-Fisher and $D_n-\sigma$ distances, spread nonuniformly within
a sphere of $\sim\!80\hmpc$ about the Local Group.
The error per galaxy is $15-21\%$ of its distance.
The galaxies are grouped into $\sim \!1200$ objects in order to 
minimize Malmquist bias.  In both cases, it is assumed that the input 
peculiar velocities are unbiased tracers of the underlying velocity 
field and are properly corrected for Malmquist bias and other 
systematics (Willick \etal 1997; Dekel \etal 1999).  We also assume 
that the peculiar-velocity errors, $\sigma_i$, are random Gaussian 
variables, independent of the underlying field.  For our current 
purpose of large-scale decomposition, both the POTENT and the WF 
analyses are applied within the framework of linear theory.

%-----------------------
\subsection{POTENT}

The latest version of the POTENT method
(originally due to Bertschinger \& Dekel 1989; Dekel, Bertschinger
\& Faber 1990) is described and evaluated in detail in Dekel \etal (1999).
It utilizes the irrotationality of linear gravitational flows,
where the velocity field is a gradient of a scalar potential,
$\vv(\vr)\!=\!-\vnabla\Phi(\vr)$. The potential field is
evaluated by integration of a smoothed version of the observed
radial velocities along radial rays from the observer,
\be
\Phi(\vr) = -\int_0^r u (r',\theta,\phi)\, dr' \ .
\label{eq:pot}
\ee
The smoothing of the radial velocities from noisy data that are sampled
nonuniformly is a nontrivial procedure which involves systematic errors.
It is done by fitting the individual velocities to a local linear field
model, with a weighting scheme that is designed for an optimal recovery
of the equal-volume smoothed field with minimum variance and minimum bias.
The smoothing is done here with a fixed Gaussian window of radius
$12\hmpc$.
Unlike the conventional nonlinear POTENT application,
the density field is derived here from the velocity field using the
linear Poisson equation, $\delta = -f(\Omega)\,\divv$.

The errors in the POTENT reconstruction are evaluated by detailed Monte Carlo
mock catalogs based on an N-body simulation by Kolatt \etal (1996).
The initial conditions of these simulations were designed using constrained
realizations to mimic the actual structure in the local universe,
but they fail to mimic the real tidal field. In order to
use the simulations for estimating the random and systematic errors
in the components of the tidal field, we therefore added to the
simulation an artificial tidal component including a bulk velocity
and a shear tensor that roughly match the tidal component recovered
from the real data.

%-------------------------------
\subsection{Wiener Filter}

The WF has been used to reconstruct the large-scale structure and the 
cosmic microwave background (CMB) temperature anisotropies from a 
variety of cosmological surveys and probes (see Zaroubi \etal 1995 and 
references therein), and in particular has been applied to the \m3\ 
velocities by Zaroubi, Hoffman and Dekel (1999).  The WF method 
implements a Bayesian approach using an assumed power spectrum 
($P(k)$) as a prior input.  Within the framework of Gaussian random 
fields it provides an estimator of the underlying field which 
coincides with the most probable field given the data, the mean field 
given the data and the maximum-entropy solution.  Yet the WF provides 
a minimum-variance estimator of the underlying field independent of 
the assumption of Gaussian random fields.  The linear theory of 
gravitational instability provides a straightforward relation between 
the fluctuation fields of velocity and density, which allows us to 
apply the WF to observed radial velocities and recover the underlying 
density field.

The WF reconstructed velocity field is given by
\be \vv _{\sss \rm WF}
(\vr) = \la \vv (\vr)\,u_i \ra\, R^{-1}_{ij} u_j \ .
\label{eq:wf}
\ee 
The correlation matrix is composed of signal plus noise, $R_{ij} = 
\la u_i u_j \ra + \delta _{ij}\sigma{^2_i}$, with $\la u_i u_j \ra $ 
the two-point radial-velocity correlation function.  The 
cross-correlation term $\la \vv _\mu (\vr)\,u_i \ra$ is calculated 
from the two-point velocity correlation tensor (G\'orski 1988 and 
Zaroubi, Hoffman and Dekel 1999).  The WF mass-density fluctuation 
field $\delta _{\sss \rm WF} $ is given by an analogous expression to 
\equ{wf} in which the first term is replaced by the cross-correlation 
matrix of density and radial velocity.  For our purpose here we do not 
use the high-resolution capability of the WF method in regions of 
high-quality data.

The uncertainties in the estimated fields are being evaluated by
constrained realizations (CRs, Hoffman \& Ribak 1991), provided that
the fields are random Gaussian fields.
First, we generate a random realization of the velocity field,
$\tilde{\vv }(\vr )$, statistically consistent with
the assumed power spectrum and the data via linear theory.
Then, we obtain a mock sample, $\tilde{u_i}$, by ``observing" this random
field at the positions of the sample galaxies.
The CR velocity field is then,
\be
\vv _{\sss \rm CR} (\vr) = \tilde{\vv }(\vr )
+  \la \vv (\vr)\,u_i \ra\, R^{-1}_{ij} (u_j - \tilde{u_j}) \ .
\ee
Constrained realizations of the density field are generated in an analogous
way.  The average of the CRs is the WF field.  Their scatter about the WF
field is the uncertainty that we attribute to the field.
This uncertainty can alternatively be evaluated from the
residual covariance matrix, but the use of CRs is preferred here,
especially for the purpose of evaluating errors in quantities
that are computed later from the WF fields.
The prior power spectrum used here is  the most likely CDM-like power
spectrum derived from the \m3\ data by Zaroubi \etal (1997),
a CDM model with $\Omega = 1$ and $h=0.75$, with tensor fluctuations
and a tilt of $n=0.81$. This large-scale tilt makes it similar to the
power-spectra of other successful cosmological model such as $\tau$CDM
and $\Lambda$CDM.

\smallskip

The two methods of reconstruction are complementary.
An advantage of POTENT is that it uses the potential-flow nature of
gravitating velocity fields, and is independent of any prior model.
Another unique feature of POTENT is the attempt to enforce uniform smoothing
throughout the volume, but this comes at the expense of amplifying
noisy features at large distances, where the sampling is poor
and the distance errors are large.
The WF reconstruction is designed to treat the random errors in
an optimal way, thus stressing only robust structures and avoiding
fake features, but this necessarily leads to nonuniform smoothing
due to artificial attenuation of the amplitude of the fields in
the noisy, outermost regions.

%%%%%%%%%%%%%%%%%%%%%%%%%%%%%%
\section{DECOMPOSITION OF THE VELOCITY FIELD}
\label{sec:decomp}

Within the linear theory of gravitational instability the velocity
field is proportional to the gravitational field, and the Poisson equation
then relates locally the fluctuation fields of density and velocity
divergent,
\be
\nabla\cdot\vv = - H_0 f(\Omega) \delta ,
\label{eq:Poisson}
\ee
where $f(\Omega) \approx \Omega^{0.6}$ (Peebles 1980).
The velocity field at a point is obtained by integration over the whole
space:
\be
\vv(\vr)=-{H_0 f(\Omega)\over 4\pi}   \int \delta (\vr')\,
{\vr'-\vr \over | \vr'-\vr |^3 }\, d^3r' \ .
\label{eq:integ}
\ee

The decomposition is done within the context of a given survey of 
peculiar velocities from which the density and 3D velocity fields are 
reconstructed in a given volume.  It is defined in reference to a 
given closed boundary about an assumed location, which separates space 
into inner and outer volumes.  The divergent component is defined by 
the integral of the density fluctuation field $\delta$ over the inner 
volume, where $\delta$ is the one reconstructed from the velocity 
survey.  The tidal component is in principle the integral over the 
outer volume, and is computed in practice as the residual after 
subtracting the divergent component from the full velocity field.  
These components relate to the particular and homogeneous solutions of 
the Poisson equation.

We choose the reference volume to be the sphere of radius $60\hmpc$
about the Local Group.  The full velocity and density fields, smoothed
%YH Both has been omitted
with a Gaussian window of radius $12\hmpc$, are reconstructed
within this volume independently  by POTENT and the WF on a cubic
grid of spacing $5\hmpc$.  %YH
The reconstructed density field is embedded
in a $128^{3}$ periodic grid, where $\delta$ is assumed to vanish
outside the reference volume.  The Poisson equation is then solved by
FFT, yielding the divergent velocity field at each point within the
inner volume.  The tidal component at each point is obtained by
subtracting the divergent component from the total velocity field.

The assumed value of $\Omega$ in the derivation of $\delta$ can be
arbitrary because it drops out in the inverse integral for the velocity
field. The value of $H_0$ also drops out by measuring distances
in units of velocity.
Note however that the POTENT errors are estimated based on a simulation
with $\Omega=1$, and that the power-spectrum prior entering the WF
reconstruction does depend on $\Omega$ and $h$.

\begin{figure}[t!]
\vspace{8.2truecm}
%\special{psfile="/home/alf2/eldar/tidal_field_apj/fig1.ps"
\includegraphics{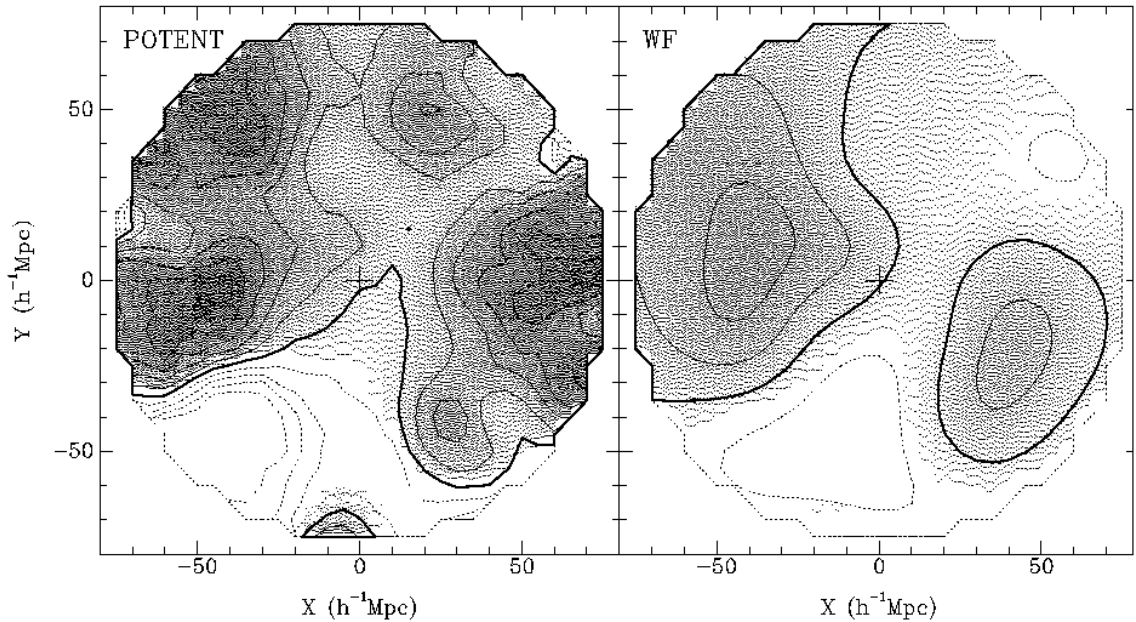}
\caption{\protect \capt
The density fields in the Supergalactic plane, recovered by POTENT (left)
and WF (right) from the \m3\ catalog of peculiar velocities.
The formal Gaussian smoothing scale is $12\hmpc$.
The Local Group is at the center.
The heavy contour marks the mean density, $\delta=0$, the dotted lines
mark negative contours, and the contour spacing is $\Delta\delta=0.2$.
Apparent are the Great Attractor on the left, the Perseus-Pisces
attractor on the right, and the void in between.
}
\label{fig:delta}
\end{figure}

\Fig{delta} illustrates the recovered density fields, by POTENT and by
the WF,
in the Supergalactic plane, extending out to $80\hmpc$.
Note the noisy features in the POTENT reconstruction at large distances,
and the attenuation of the signal in the WF reconstruction.
Both maps show the dominant structures of the Great Attractor (GA)
and Perseus-Pisces (PP), separated by a large underdense region.

\begin{figure}[t!]
\vspace{12.0truecm}

%\special{psfile="/home/alf2/eldar/tidal_field_apj/fig3.ps"
\includegraphics{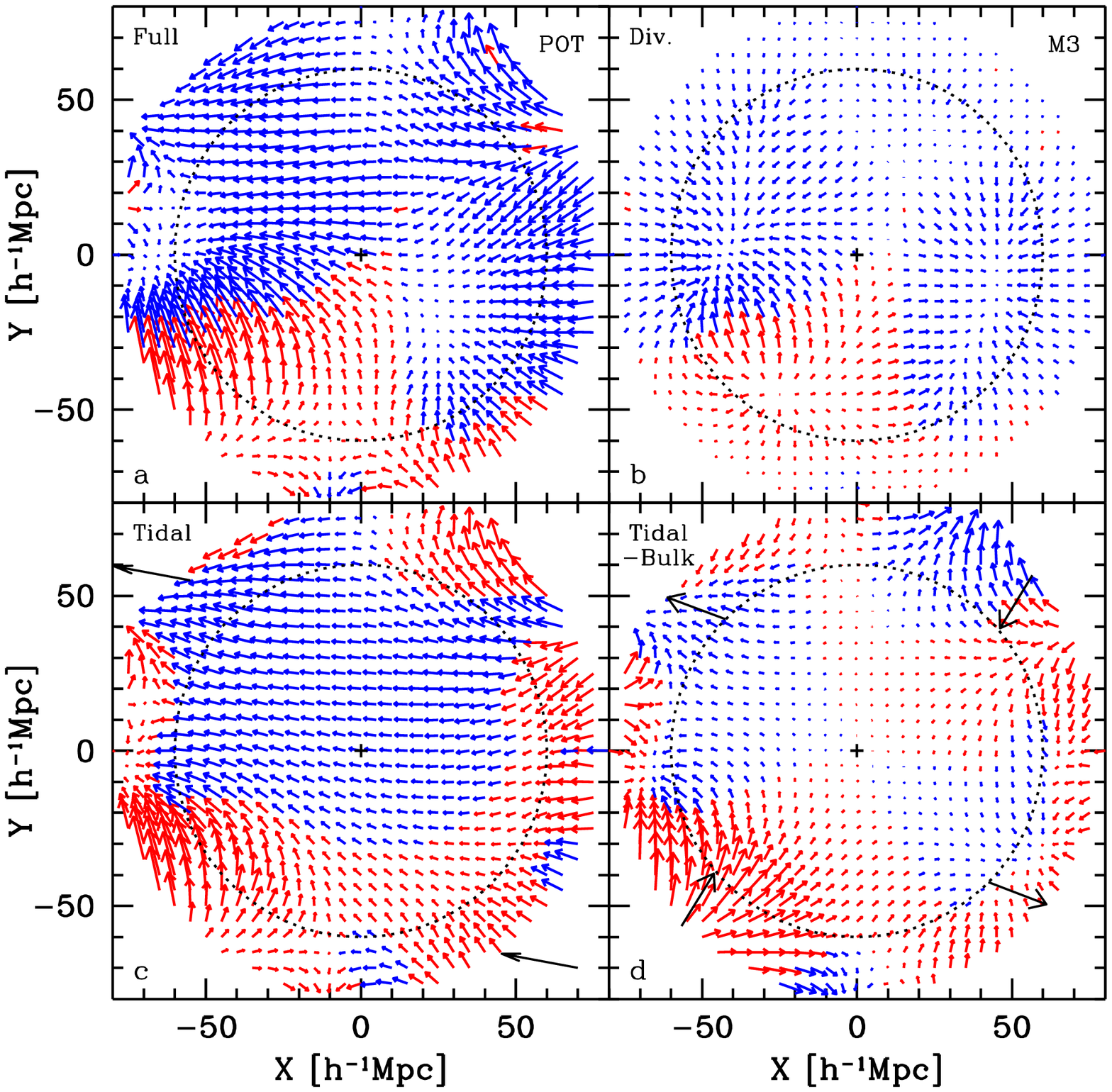}

\caption{\protect \capt
Decomposition of the POTENT velocity field in the Supergalactic plane,
with respect to the sphere of $60\hmpc$ about the Local Group (center).
The Velocities are measured in $\!\hmpc$ ($1\hmpc=100\kms$).
(a) The full velocity field.
(b) The divergent component due to the mass fluctuations within
    the sphere (shown in \Fig{delta}.
(c) The tidal component due to the mass distribution outside the sphere.
(d) The residual after subtracting the bulk velocity from the tidal component,
    including quadrupole and higher moments.
The black long arrows in the bottom panels show the projected directions
of the bulk velocity and two of the shear eigenvectors respectively.
(The red and blue in the upper panels correspond to positive and negative
$\divv$ respectively.  In the bottom-left panel they distinguish between
velocities according to whether the angle they form with the direction of
the bulk velocity (long arrows) is larger or smaller than $30^\circ$.
In the bottom-right panel the colors refer to the direction of the
eigenvector with the largest eigenvalue.)
}
\label{fig:v-pot}
\end{figure}

\begin{figure}[t!]
\vspace{12.0truecm}

%\special{psfile="/home/alf2/eldar/tidal_field_apj/fig2.ps"
\includegraphics{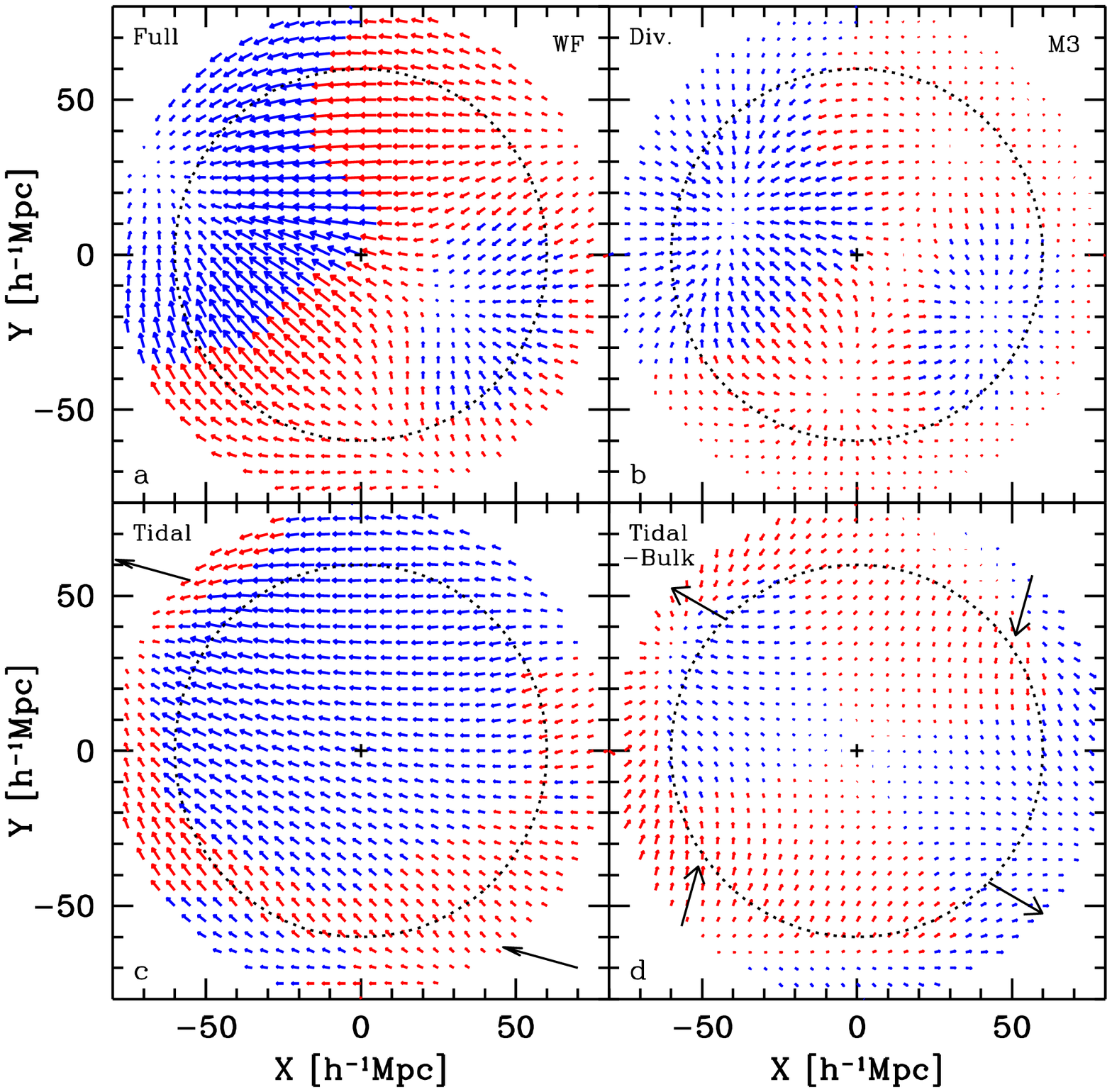}

\caption{\protect \capt
Same as \Fig{v-pot} but for the WF reconstruction.
}
\label{fig:v-wf}
\end{figure}

The corresponding velocity fields, in the Supergalactic plane
and projected onto it, 
are shown in the top-left panels of \Fig{v-pot} and \Fig{v-wf}.
Two other panels in these figures show the divergent and tidal
components.
The divergent fields are dominated by the two characteristic infall patterns
into the two attractors, GA and PP, and the outflows from the voids
in between. A bulk flow, if it exists, does not seem to be
a dominant feature of the divergent component.
The tidal field, on the other hand, is clearly dominated by a
coherent bulk velocity.

%%%%%%%%%%%%%%%%%%%%%%%%%%%%%%
\section{TIDAL FIELD: BULK VELOCITY AND SHEAR}
\label{sec:tidal}

To analyze the tidal field in more detail, it is further decomposed
into several components by fitting a linear model of the sort
\be
v_\alpha(\vr) = B_\alpha + (\tilde{H}\, \delta_{\alpha\beta} +
\Sigma_{\alpha\beta}) r_\beta \ .
\label{eq:shear}
\ee
Here, the indices $\alpha$ and $\beta$ refer to the 3 Cartesian coordinates,
$B_\alpha$ is the bulk velocity,
$\tilde{H} \delta_{\alpha\beta}$ accounts for a possible
local isotropic perturbation about the Hubble expansion
($\delta_{\alpha\beta}$ is the identity matrix),
and $\Sigma_{\alpha\beta}$ is the traceless shear tensor.
In the terminology of potential theory, the bulk and shear terms are
the dipole and quadrupole respectively. The expansion does not
contain an antisymmetric part because the flow is assumed to be
irrotational.
The fit is done by an unweighted volume average over the grid points inside
the reference volume.
In the case of the tidal field the isotropic term,
$\tilde{H} \delta_{\alpha\beta}$, vanishes by construction, because it
corresponds to the divergent component of the flow.

The traceless shear tensor is then diagonalized, and we denote its eigenvalues
in decreasing order $\lambda_1>\lambda_2>\lambda_3$, where
$\lambda_1+\lambda_2+\lambda_3=0$.
The eigenvalues $\lambda_1$ and $\lambda_3$ correspond to the dilational
and compressional modes respectively.

The fourth panels of \Fig{v-pot} and \Fig{v-wf} show the result of
subtracting the bulk velocity from the tidal field. This field is
dominated by a clear quadrupole pattern, for which the Supergalactic
plane is close to a principal plane. The eigenvector of the largest
eigenvalue lies along the line from bottom-right to top-left.

%----------------------------
\subsection{Bulk Velocity of the Tidal Field}

\begin{figure}[t!]
\vspace{8.0truecm}

%\special{psfile="/home/alf2/eldar/tidal_field_apj/fig4.ps"
\includegraphics{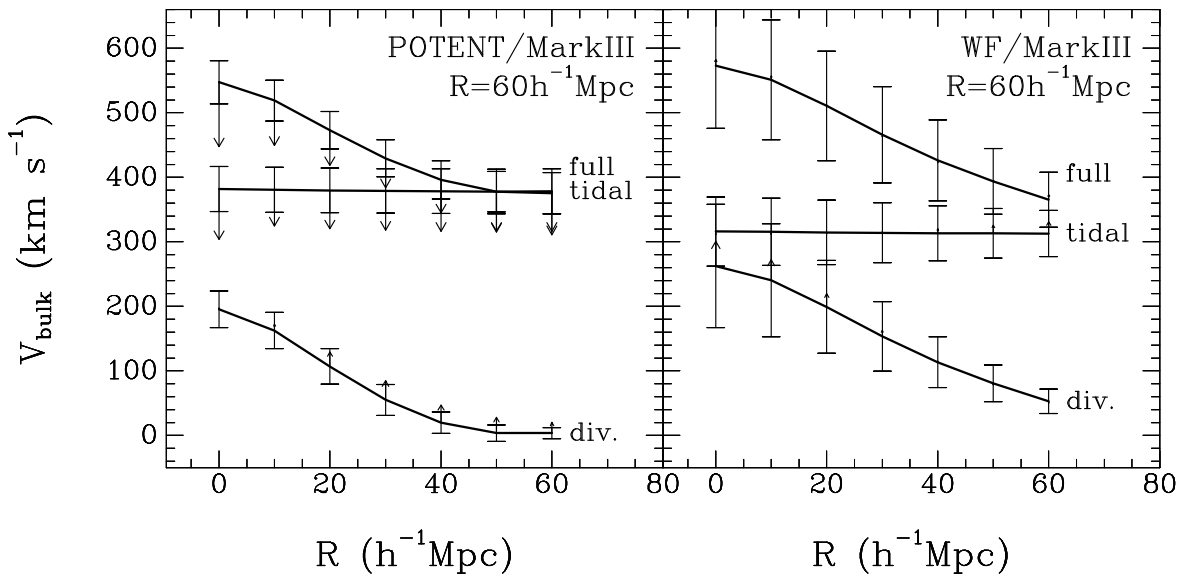}

\caption{\protect \capt 
The amplitude of bulk velocity in the CMB 
frame in spheres of radius $R$ about the Local Group.  It is derived 
by equal-volume vector averaging of the POTENT (left) and WF (right) 
velocity fields.  Shown are the total bulk velocity and its two 
components, divergent and tidal.  The error bars mark random errors, 
and the arrows mark crude estimates of systematic errors (POTENT 
only).  }
\label{fig:bulk}
\end{figure}

\Fig{bulk} shows the amplitude of the bulk velocity in the CMB frame, 
averaged in spheres of radius $R$ about the Local Group.  Shown are 
the total bulk velocity and its two components, divergent and tidal.  
The divergent contribution is significant at small radii, indicating 
that local density structures like GA and PP are responsible for up to 
one half of the Local-Group velocity.  The divergent bulk velocity 
tends to zero towards $R=60\hmpc$.  Note that this is not guaranteed a 
priori; a nonzero bulk flow can in principle be associated with local 
perturbations (e.g., in the case of a dominant attractor near the edge 
of the volume).  The vanishing divergent bulk velocity reflects the 
general mirror symmetry between GA and PP.

The tidal bulk velocity from POTENT is $366 \pm 125\kms$ in the direction
$(L,B)=(167^\circ,\, -10^\circ)$, and from WF it is
$309 \pm 125\kms$ in the direction $(L,B)=(165^\circ,\, -15^\circ)$.
The two methods yield very similar directions that differ by only $8^\circ$.
The amplitudes are consistent with each other; they differ in the
direction expected due to the differences between the
methods as explained in \se{recon}.
The two tidal bulk velocities form an angle of $\sim 34^\circ $ and
$\sim 26^\circ $ respectively with the Local Group motion in the
CMB frame, implying, again, that about half of this Local-Group velocity, of
$627 \pm 22 \kms$ (Lineweaver \etal 1996),
is induced by density structures on scales beyond $60\hmpc$.

\begin{figure}[t!]
\vspace{17.0truecm}

%\special{psfile="/home/alf2/eldar/tidal_field_apj/fig6.ps"
\includegraphics{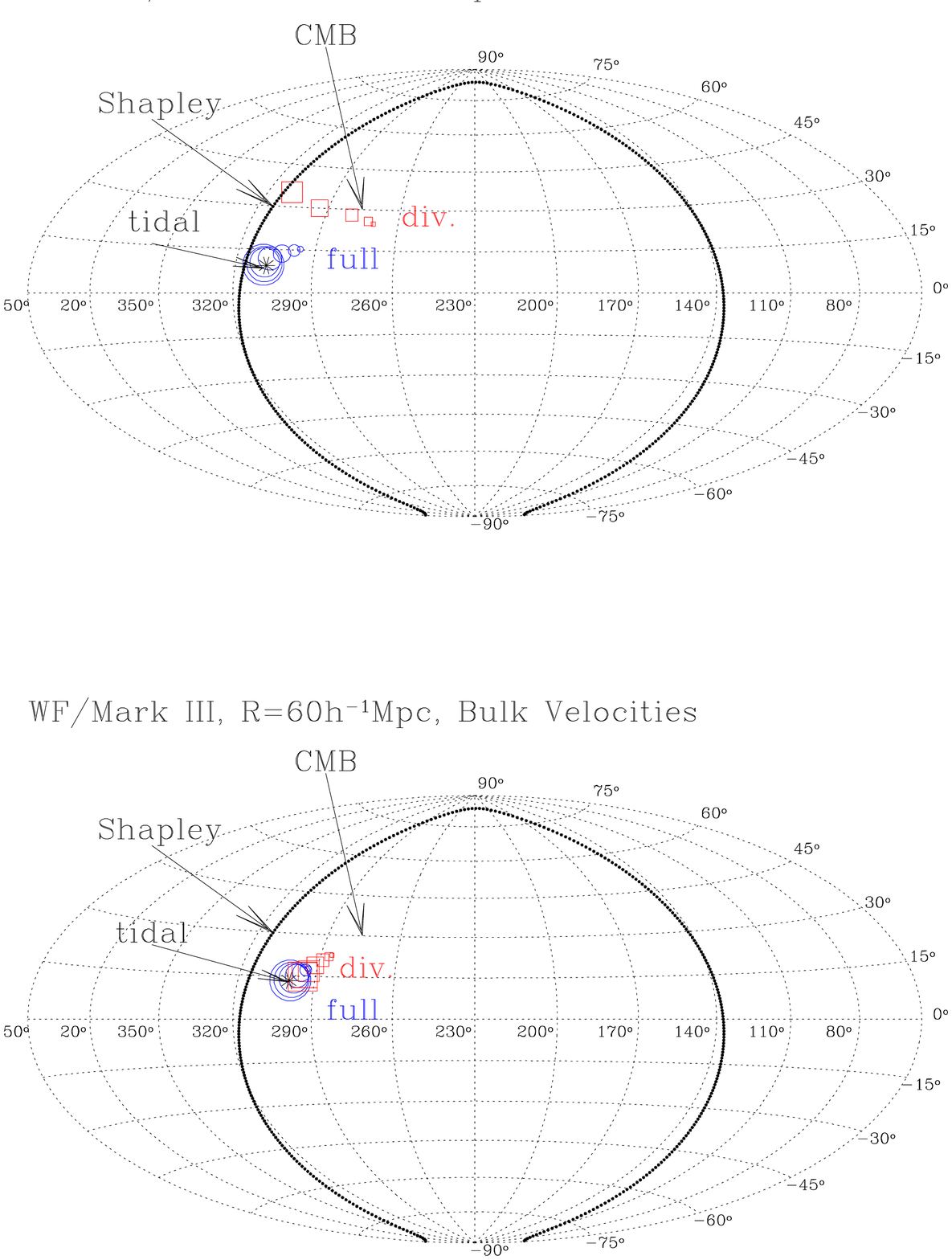}

\caption{\protect \capt
The directions of the full bulk velocity and its divergent and
tidal components, in Galactic coordinates.
The bulk velocities are derived inside spheres of radii $20, 30, ... 60\hmpc$,
and the circle size is proportional to the sphere radius.
}
\label{fig:aitoff_converge}
\end{figure}

\Fig{aitoff_converge} shows the direction of the bulk velocity
as a function of sphere radius, for the full velocity field,
and for its two components.
The WF reconstruction reveals a very robust direction for the bulk
velocity, as expected,
while the directions in the POTENT case fluctuate a bit more due
to the random errors at radii larger than $40\hmpc$.

%-----------------------------------------
\subsection{Shear Tensor of the Tidal Field}

As seen in \Fig{v-pot} and \Fig{v-wf},
the residual tidal field after the bulk velocity is subtracted out
is dominated by a quadrupole whose dilational eigenvector is
roughly aligned with the direction to the Shapley Concentration.

The one-dimensional rms residual about the linear fit,
($\sigma_{\rm v} $),
which is almost the same in the three directions, is
$\sigma_{\rm v} \simeq 145 \kms$ for POTENT and
$\simeq 28 \kms$ for WF. The
eigenvalues for the POTENT field are $0.037\pm 0.029$, $0.008\pm
0.032$, and $-0.045\pm 0.027$.  For the WF field they are quite
consistent: $0.036\pm 0.007$, $-0.003\pm 0.009$, and $-0.033\pm
0.007$.

\begin{figure}[t!]
\vspace{17.0truecm}
%\special{psfile="/home/alf2/eldar/tidal_field_apj/fig5.ps"
\includegraphics{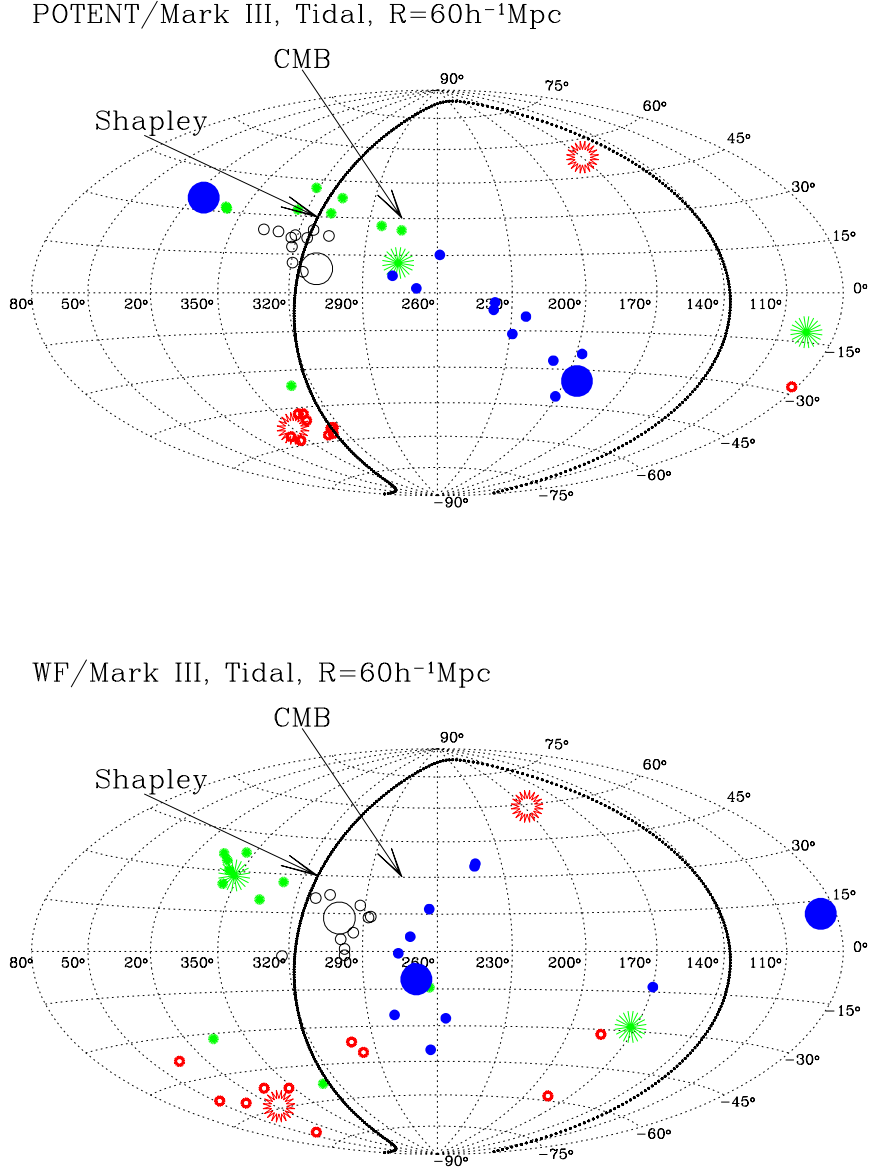}
\caption{\protect \capt
The directions of the tidal bulk velocity
(open circles), and the three eigenvectors of the shear:
the dilation (skeletals), the compression (starred), and the
middle-eigenvalue vector (solid).
The large symbols are for the real data, and the corresponding small symbols
are for 10 random realizations generated by mock catalogs for the POTENT
and by CRs for WF.
Also shown are the anti-apex of the three eigenvectors.
}
\label{fig:aitoff}
\end{figure}

\Fig{aitoff} shows the directions of the bulk velocity
and the three shear eigenvectors of the tidal field.
The uncertainties about these directions are evaluated
by applying the tidal field decomposition to 10 random mock catalogs
(POTENT) and CRs (WF). The resultant spread in directions
is shown in the figure.

Although the dilational eigenvectors as obtained from the two
reconstruction methods are about $70 ^\circ$ apart, given the
uncertainties they are both not that far from the direction
of the tidal bulk velocity (deviations of $\sim 30 ^\circ$ for POTENT
and $\sim 50 ^\circ$ for EF), and they are both in the crude
general direction of the Shapley concentration.
The corresponding compressional eigenvectors are only
about $17^\circ$ from each other, while the middle eigenvectors
are very uncertain and are quite far apart.

%-------------------------------------
\subsection{Sensitivity of WF to the Prior Power Spectrum}
\label{sec:pk}

We examine here the sensitivity of our WF tidal results
to the assumed prior power-spectrum.
The maximum-likelihood determination of the power-spectrum
parameters (Zaroubi \etal 1997),
for tilted, flat $\Lambda$CDM, COBE normalized with tensor fluctuations,
put $1\sigma$ constraints on a degenerate combination of
the parameters, roughly $\Omega h_{50}^{1.3} n^{3.4} = 0.8\pm 0.12$
(where $h_{50} \equiv H_0/50\kmsmpc$).

%%%%% [Table 1}
\begin{deluxetable}{lcccccccc}
\def\deg{^\circ}
%\label{tab:}
%\scriptsize
\tablewidth{0pt}
\tablecaption{Sensitivity of the Tidal Bulk and Shear to the Prior Power
Spectrum}
\tablehead{
$\Delta (\Omega h^{1.3}n^{3.4})$ & $\Omega $ & $h$ & $n$ &
 $B_t (\kms)$ & $(L,B)$ & $\lambda _1$ & $\lambda _2$ & $\lambda _3$
}
\startdata
{\bf 0} & {\bf 1} & {\bf 0.75} & {\bf 0.81} &
{\bf 309} &{\bf $(164.5\deg ,-15.0\deg)$} &{\bf 0.036} &{\bf-0.033} &
{\bf -0.003} \nl \nl
 0 & 0.5 & 0.75 & 0.993 &
 343 &$(165.2\deg ,  -15.1\deg)$ & 0.048 & -0.041 & -0.007 \nl
 0 & 1 & 0.60 & 0.882 &
 322 & $(164.7\deg ,  -15.2\deg)$ & 0.040 & -0.036 & -0.004 \nl
 -0.12 & 0.855 & 0.75 & 0.81 &
 315 & $(165.3\deg ,  -14.9\deg)$ & 0.038 & -0.034 & -0.004 \nl
 -0.12 & 1 & 0.75 & 0.7735 &
 305 & $(165.0\deg ,  -14.8\deg)$ & 0.034 & -0.031 & -0.003 \nl
 -0.12 & 1 & 0.665 & 0.81 &
 313 & $(165.2\deg ,  -14.9\deg)$ & 0.037 & -0.033 & -0.004 \nl
 +0.12 & 1    & 0.75 & 0.843 &
 311 & $(164.0\deg ,  -15.2\deg)$ & 0.037 & -0.034 & -0.003 \nl

\enddata
\end{deluxetable}

In Table 1 we present the results extracted from several WF
reconstructions using different values
for $\Omega $, $h_{50}$ and $n$ that obey the maximum-likelihood constraint,
as well as various combinations of these parameters that lie on the
$1\sigma$ likelihood contour.  The table displays the resultant
tidal-field quantities of bulk velocity and shear.
The variations in the bulk velocity are less than $10\%$ in magnitude and
$1^\circ$ in direction. The variations in the eigenvalues of the shear
tensor are typically $10-30\%$. The WF results are thus quite robust.

%-----------------------------------------
\subsection{Preliminary Results from SFI}

The method described in this paper is being applied to other data sets
and compared to complementary information from redshift surveys
(in preparation).
To have a feeling for the robustness of our obtained tidal field,
we briefly describe here preliminary results from the SFI catalog
of peculiar velocities (Haynes \etal 1999a,b). This catalog contains
Tully-Fisher distances for $\sim 1300$ Sc galaxies. Its spatial extent
is similar to the \m3\ sample, but the sampling is sparser and
more uniform throughout the volume.
Malmquist bias has been corrected in SFI as explained in
Freudling \etal (1999).

\begin{figure}[t!]
\vspace{12.0truecm}

%\special{psfile="/home/alf2/eldar/tidal_field_apj/fig7.ps"
\includegraphics{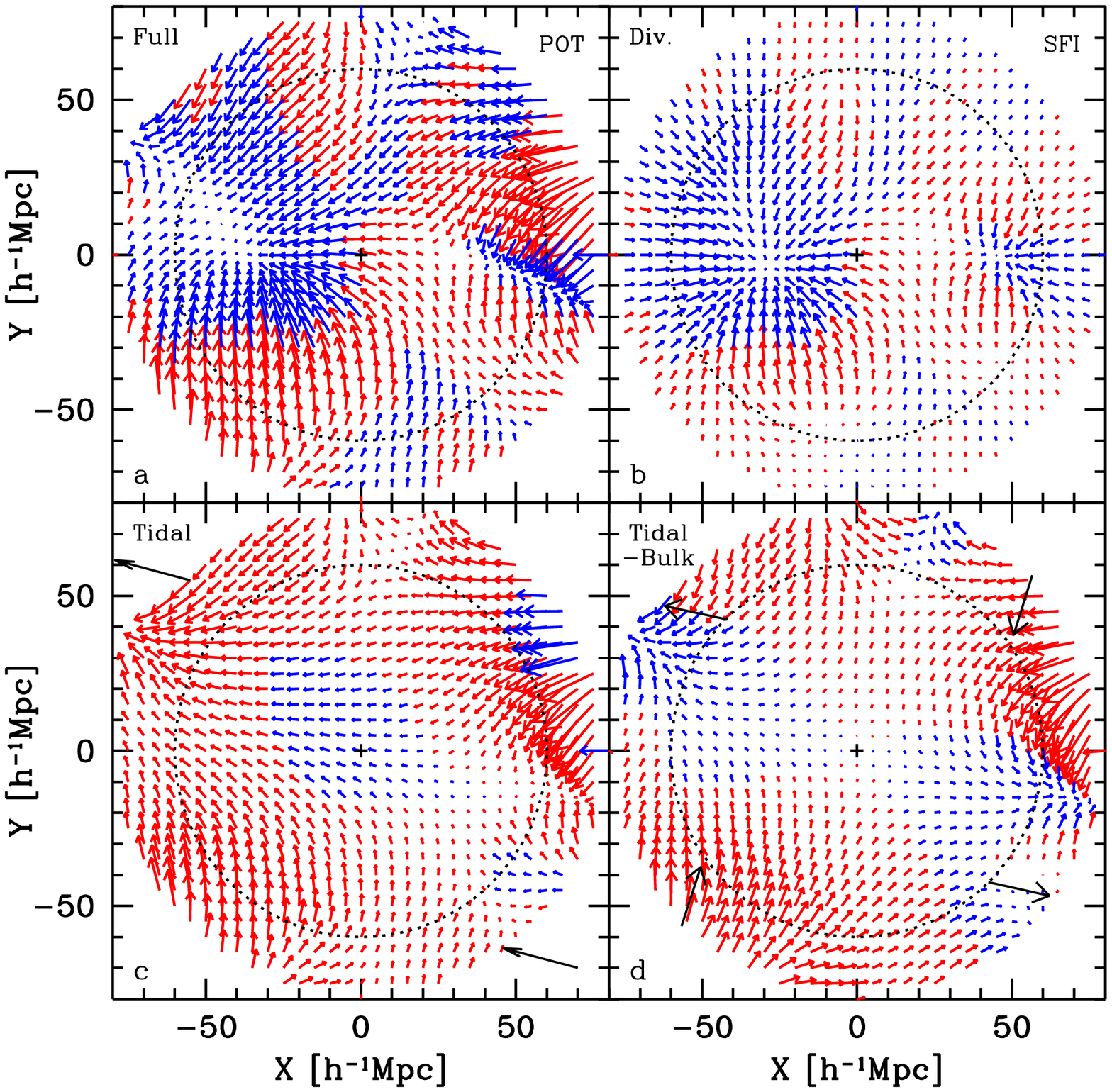}
% angle=0 hscale=60 vscale=60 hoffset=0  voffset=-200}

\caption{\protect \capt
Same as \Fig{v-pot}, but for the SFI data.
}
\label{fig:sfi}
\end{figure}

\Fig{sfi} shows the decomposition of the POTENT velocity field
from the SFI data, similar to \Fig{v-pot} for \m3.
Note the similarity between the recovered tidal fields from the two
data sets.  The best-fit tidal bulk velocity of SFI is $255\kms$ towards
$(L,B)=(165,\, -17)$, which is fully consistent with that of \m3.
The directions of the shear eigenvectors are also very similar for
the two data sets, with SFI eigenvalues $(0.068,\, 0.008,\, -0.076)$,
repeating the pattern of two eigenvalues with comparable absolute
values. Despite the large uncertainties, the tidal field is surprisingly
robust.

%%%%%%%%%%%%%%%%%%%%%%%%%%%%%
\section{THE SHAPELY ATTRACTOR}
\label{sec:shapley}

The components of the shear tensor provide an indication 
for the source of the tidal field, which we saw is responsible
for about one half of the Local Group velocity.
Since the bulk velocity and the dilational eigenvector point in
the same general direction, it may be useful to fit the tidal
field with a toy model of a single attractor --- a point-mass
or a sphere of mass excess $\delta M$ centered at a distance
$D$ from the Local Group.
Within that model the distance is estimated by $D \sim 2 B/\lambda_1$.
Given the values obtained above from the Mark III data we estimate
\be
D \sim (150 - 200) \hmpc \ .
\ee
In linear theory this fluctuation generates a velocity field
\be
v(\vr) = {2f(\Omega)\over 3 \Omega}{G \, \delta M\over H_0 D^2} \ .
\label{eq:deltaM}
\ee
Substituting the estimated value of $D$, we obtain
\be
\delta M \sim (2-5)\times 10^{17} M_\odot h^{-1} \Omega^{0.4} . 
\ee

This single attractor model closely coincides with the Shapley
Concentration of rich galaxy clusters.
Reisenegger \etal (2000), who studied the dynamics of the Shapley
Concentration, obtained using a spherical
infall model an upper limit of $M < 1.3\times 10^{16} M_\odot h^{-1}$
(for an Einstein-de Sitter universe) and
$M < 8.5\times 10^{15} M_\odot h^{-1}$ (for an empty universe),
consistent with the predictions of our toy model.
They located the Shapley supercluster at a redshift $z \approx 0.048$,
close to our estimate of $D$.  The direction of the Shapley Concentration
lies close to the directions of the bulk velocity and the dilational
eigenmode of the tidal field, and almost in between.
These all point to the important role of the Shapley Concentration
in inducing the tidal field within the local sphere of $60 \hmpc$.

However, this indication should not be overinterpreted.  The single
attractor model is clearly an oversimplification which provides only a
limited fit to the data.  This is manifested by the eigenvalue
ratio of $\lambda_3/\lambda_1 \simeq -1$ for the real tidal shear tensor,
compared to the predicted value of $-0.5$ of the single-attractor model.
Shapley is likely to be an important tidal source, but it is certainly
not the only relevant external perturbation.

A better fit to the tidal field is provided by a slightly more complicated 
toy model which includes one point-mass attractor and two negative point
masses representing voids of density below the cosmological average. 
We fix the attractor position roughly at the Shapely Concentration, 
at a distance of $150\hmpc$ in the Supergalactic direction 
$(L,B)=(142^\circ,0^\circ)$, 
and allow the associated mass to be a free parameter. Two voids are allowed, 
with free masses and free positions, limited to lie outside the volume 
boundary of $60\hmpc$ and inside $100\hmpc$. This toy model has 9 free 
independent parameters, while there are 8 
constraints imposed by the tidal bulk velocity and shear tensor.  
A perfect solution is not guaranteed even when the number of parameters
matches the number of constraints because the system of equations is 
nonlinear, and because we limit the search volume to $100\hmpc$. 
We determine the errors for each component using mock catalogs based on
simulations, and determine the best-fit parameters by minimizing the sum 
of residuals over the 8 components:
\be
\sum _{\alpha } {(B^{\rm model}_{\alpha } - B^{\rm obs}_\alpha)^2\over
\sigma _{B_\alpha}^2} 
+ \sum _{\alpha, \beta } {(\Sigma^{\rm model}_{\alpha \beta } -
\Sigma^{\rm obs}_{\alpha \beta})^2\over \sigma _{\Sigma_{\alpha \beta}}^2} \ .
\ee
The model bulk velocity is the sum of the contributions of each of the three
point masses, of the sort 
%YH The numerical factor in  the
%next 2 eqs is 2/3 and NOT 3/3
$2f(\Omega)G/(3\Omega H_0)\, \delta M\, r_\alpha /r^3$,     %YH
and the model shear is similarly a sum of the three contributions of the sort
$2f(\Omega)G/(3\Omega H_0^2)\, \delta M\,                       %YH
(3\hat{r}_\alpha \hat{r}_\beta-\delta _{\alpha \beta}) / r^3$.

\begin{figure}[t!]
\vspace{12.5truecm}

\includegraphics{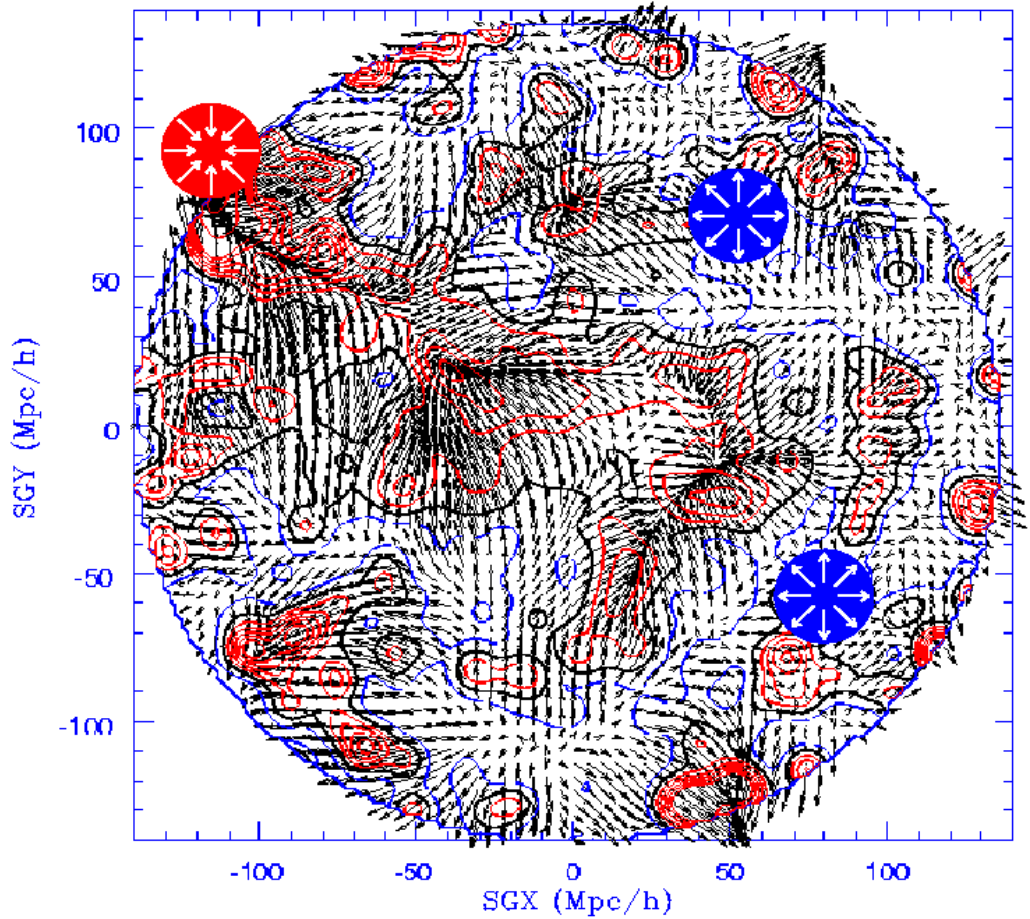}

\caption{\protect \capt
A 3-mass toy model on top of the PSCz galaxy distribution in the 
Supergalactic Plane.
Marked are the positions of the attractor and two voids, in the toy model that
best fits the tidal field as derived by POTENT from \m3. 
Shown at the background are the density and velocity fields from the PSCz 
redshift survey of IRAS galaxies. The Local Group is at the center.
The structures towards the top-left are the Great Attractor in the foreground
and the Shapely Concentration in the background.
The structure to the opposite direction is Perseus Pisces.
The Come Supercluster is at a distance of $\sim 70-80\hmpc$ towards the top.
Ursa Major is a nearby cluster towards top-right.
}
\label{fig:voids}
\end{figure}

The best toy model we found
has an attractor mass of $1.6\times 10^{17} M_\odot h^{-1}$,
and effective masses of $-0.5\times 10^{17} M_\odot h^{-1}$ associated 
with each of the two voids, located at $(D,L,B)=(88\hmpc,54^\circ,0^\circ)$ 
and $(100\hmpc,324^\circ,11^\circ)$ 
respectively. Note how close these structures are to the Supergalactic Plane.
\Fig{voids} shows the positions of these three elements of the toy model 
on top of the map of the density and velocity fields in the Supergalactic Plane
as derived from the PSCz redshift survey of IRAS galaxies (Schmoldt 
\etal\ 1999).
One void lies roughly along the principal mode of dilation, namely along 
the line connecting the Local Group with Shapely but to the opposite 
direction, behind the Perseus Pisces (PP) supercluster. 
The other void lies in an orthogonal 
direction, behind the nearby Ursa Major (UM) cluster.
The tidal bulk velocity induced by this toy model is only about 
$20^\circ$ away from the real bulk velocity by POTENT from \m3.
The shear eigenvalues of the toy model are $(0.029,0.012,-0.041)$,
compared to the real eigenvalues $(0.037,0.008,-0.045)$, and they point 
in directions that are less than $15^\circ$ away from the real directions. 
This toy model is thus quite successful.

Note in \Fig{voids} that the dynamical voids indicated by the tidal field 
indeed lie in the vicinity of two out of a few extended voids as seen in 
the PSCz map.
Another void is apparent along the line connecting the Local Group to the
UM void but in the opposite direction, termed the Sculptor void.
A fit of somewhat lower quality but still with the right main features
is obtained by a toy model in which the PP void is replaced by a 
void in the vicinity of Sculptor.

%%%%%%%%%%%%%%%%%%%%%%%%%%%%%
\section{CONCLUSION}
\label{sec:conc}

The main purpose of this paper was to present an algorithm for the
decomposition of the large-scale tidal velocity field.  The algorithm
is based on solving the linear Poisson equation given the
reconstructed density field inside a certain volume, while assuming
that the fluctuations vanish outside the volume.  The tidal field is
then obtained by subtracting this particular solution from the full
reconstructed velocity field.

This algorithm has been applied to the velocity and density fields
recovered from the \m3\ peculiar-velocity data by the complementary
POTENT and WF methods.
The tidal field was calculated with respect to the sphere of radius $60\hmpc$
about the Local Group.

The tidal field is characterized by a bulk velocity and a shear 
tensor.  We found that about half the amplitude of the CMB dipole is 
due to mass-density fluctuations outside the reference sphere.  The 
dilational eigenvector (with the largest eigenvalue) of the shear 
tensor is roughly aligned with the bulk velocity, which indicates the 
important dynamical role of an attractor of mass $(2-5)\times 10^{17} 
M_\odot h^{-1} \Omega^{0.4}$ at a distance of $150-200 \hmpc$, behind 
the Great Attractor, roughly coinciding with the Shapley 
Concentration.  However, the tidal field must also involve additional 
structures.  The addition of two voids in the Supergalactic Plane 
improves the fit: one void behind the Ursa Major cluster, and the 
other either behind the Perseus-Pisces supercluster or the Sculptor 
void in an orthogonal direction.  These roughly coincide with voids in 
the PSCz redshift survey.

Our conclusions are strengthened by the agreement between the results 
obtained using the two complementary reconstruction methods, which 
treat the random errors in very different ways.  The two estimators 
are likely to bracket the actual amplitude of the tidal field.

The obtained tidal field from the SFI data is consistent with that 
obtained from the \m3\ data, which is quite surprising in view of the 
expected large errors.  Consistency with other data and its 
implications are now under investigation.

The tidal field we find have implications for the issue of convergence 
of the bulk velocity as measured from different velocity surveys on 
different scales (a summary in Dekel 2000), and for the question of 
the effective depth for the recovery of the CMB dipole motion of the 
Local Group.  The latter is crucial for evaluating the cosmological 
density parameter from the CMB dipole versus the dipole from a 
redshift survey (e.g., Lahav, Kaiser and Hoffman 1990; Juszkiewicz, 
Vittorio and Wyse 1990).

%%%%%%%%%%%%%%%%%%%%%%%%%%%%%%%%%%%%%%%%%%%%%%%%%%%%%%%
\acknowledgments
%YH grant #s
This research has been supported by grants to YH and to AD by the 
US-Israel Binational Science Foundation (94-00185 and 98-00217) and 
the Israel Science Foundation (103/98 and 546/98).  YH acknowledges 
stimulating discussions with R. van de Weygaert and the hospitality of 
the Kapteyn Institute of Astronomy.

%%%%%%%%%%%%%%%%%%%%% REFERENCES %%%%%%%%%%%%%%%%%%%%%%%%%%%%%%%%%
\newpage
\def\re{\reference}

\end{document}

%%%%%%%%%%%%%%%%%%%%%%%%%%%%%%%%%%%%%%%%%%